\documentclass
[
amssymb,prl,aps,twocolumn,floatfix,tightenlines,superscriptaddress]{revtex4-1}
\pdfoutput=1
\usepackage{graphics}
\usepackage{graphicx,color}
\usepackage{subfigure}
\usepackage{float}
\usepackage[ansinew]{inputenc}
\usepackage{graphicx}
\usepackage{amsmath}
\usepackage{amssymb}
\usepackage{dsfont} 
\usepackage{hyphenat}
\usepackage[percent]{overpic}
\usepackage{xcolor}
\usepackage{appendix}

\newcommand{\ket}[1]{\left | #1 \right \rangle}
\newcommand{\bra}[1]{\left \langle #1 \right |}

\newcommand{\dist}[1]{\mathcal{#1}} 

\begin{document}

\title{On the experimental verification of quantum complexity in linear optics}
\author{Jacques Carolan}
\author{Jasmin D. A. Meinecke}
\author{Pete Shadbolt}
\author{Nicholas J. Russell}
\affiliation{Centre for Quantum Photonics, H. H. Wills Physics Laboratory \& Department of Electrical and Electronic Engineering, University of Bristol, Merchant Venturers Building, Woodland Road, Bristol, BS8 1UB, UK}
\author{Nur Ismail}
\author{Kerstin W{\"o}rhoff}
\affiliation{Integrated Optical Microsystems Group, MESA+ Institute for Nanotechnology, University of Twente, Enschede, The Netherlands}
\author{Terry Rudolph}
\affiliation{Institute for Mathematical Sciences, Imperial College London, London SW7 2BW, UK}
\author{Mark G. Thompson}
\affiliation{Centre for Quantum Photonics, H. H. Wills Physics Laboratory \& Department of Electrical and Electronic Engineering, University of Bristol, Merchant Venturers Building, Woodland Road, Bristol, BS8 1UB, UK}
\author{Jeremy L. O'Brien}
\author{Jonathan C. F. Matthews}
\email{jonathan.matthews@bristol.ac.uk}
\author{Anthony Laing}
\email{anthony.laing@bristol.ac.uk}
\affiliation{Centre for Quantum Photonics, H. H. Wills Physics Laboratory \& Department of Electrical and Electronic Engineering, University of Bristol, Merchant Venturers Building, Woodland Road, Bristol, BS8 1UB, UK}
\date{\today}

\begin{abstract}
The first quantum technologies to solve computational problems that are beyond the capabilities of classical computers are likely to be devices that exploit characteristics inherent to a particular physical system, to tackle a bespoke problem suited to those characteristics.
Evidence implies that the detection of ensembles of photons, which have propagated through a linear optical circuit, is equivalent to sampling from a probability distribution that is intractable to classical simulation.
However, it is probable that the complexity of this type of sampling problem means that its solution is classically unverifiable within a feasible number of trials, and the task of establishing correct operation becomes one of gathering sufficiently convincing circumstantial evidence.
Here, we develop scalable methods to experimentally establish correct operation for this class of sampling algorithm, which we implement with two different types of optical circuits for 3, 4, and 5 photons, on Hilbert spaces of up to $50,000$ dimensions.  With only a small number of trials, we establish a confidence $>99\%$ that we are not sampling from a uniform distribution or a classical distribution, and we demonstrate a unitary specific witness that functions robustly for small amounts of data.  Like the algorithmic operations they endorse, our methods exploit the characteristics native to the quantum system in question.  Here we observe and make an application of a ``bosonic clouding'' phenomenon, interesting in its own right, where photons are found in local groups of modes superposed across two locations.  Our broad approach is likely to be practical for all architectures for quantum technologies where formal verification methods for quantum algorithms are either intractable or unknown.
\end{abstract}

\maketitle
The construction of a universal quantum computer, capable of implementing any quantum computation or quantum simulation, is a  major long term experimental objective. However, it is expected that non-universal quantum machines, that exploit characteristics of their own physical system to solve specific problems, will out-perform classical computers in the near-term \cite{AspuruGuzik:2012ho}.
Ensembles of single photons in linear optical circuits are a recently proposed example: despite being non-interacting particles, their detection statistics are described by functions that are intractable to classical computers --- matrix permanents \cite{Scheel:2004tt}.
It is therefore believed that linear optics constitutes a platform for the efficient sampling of probability distributions that cannot be simulated by classical computers, with strong evidence provided in the case of circuits described by large random matrices \cite{Aaronson:2011tj, Broome:2013ti, Spring:2013to, Crespi:2012fu, Tillmann:2012ux}.

A universal quantum computer, running for example Shor's factoring algorithm \cite{Shor:1999ul}, creates an exponentially large probability distribution with individual peaks at highly regular intervals that facilitate the solution to the factoring problem allowing efficient classical verification, as is the case for all problems in the NP complexity class \cite{Nielsen:2011vx}.  Accordingly, correct operation of the quantum computer is confirmed.  In contrast, it is not clear that similarly useful structure exists in the exponentially large probability distribution that is sampled when photons are detected after a random transformation.  And since such boson sampling problems \cite{Aaronson:2011tj} are related to the harder $\#$P complexity class \cite{va-tcs-8-189}, it is not understood how to verify correct operation for large versions of a boson sampling machine, with formal verification likely to be classically intractable.

\begin{figure*}[t!]
\includegraphics[trim=20 30 40 10, clip, width=1\linewidth]{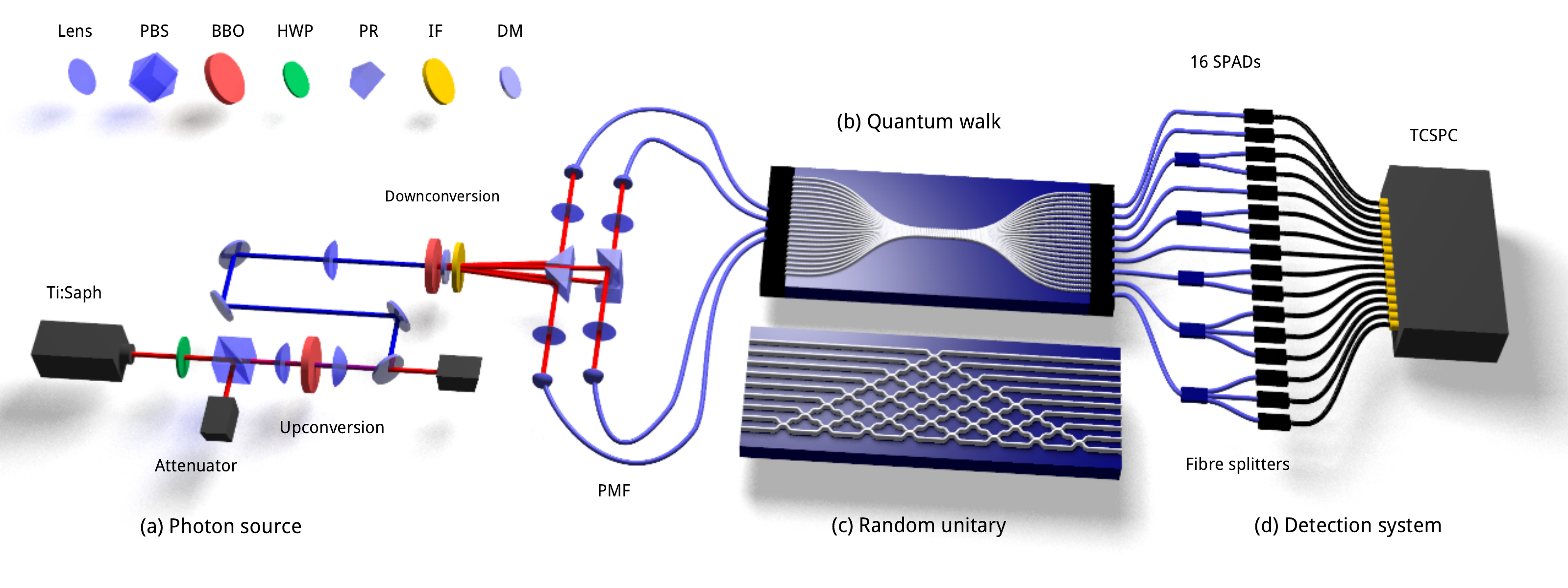}
\caption{Experimental setup to generate (a), interfere (b,c) and detect (d) single photons. (a) 780nm laser light from a 140fs pulsed Titanium:Sapphire laser was attenuated with a half wave plate (HWP) and polarising beams splitter (PBS), before frequency doubling with a type-I BBO nonlinear crystal.  The subsequent 390nm light was reflected from four dichroic mirrors (DM) and focused onto a type-I BiBO nonlinear crystal to generate double pairs of photons through spontaneous parametric down conversion.  After passing through an interference filter (IF) photons are reflected off a prism (PR) and collected into polarisation maintaining fibres (PMF) which are butt-coupled, via a V-groove fibre array, to either (b) the QW chip, or (c) the RU chip. Outgoing photons are coupled from the chip using a second fibre array, either directly to 16 single photon avalanche diodes (SPADs) (d), or via a network of fibre splitters. Detection events are time-correlated and counted using a 16-channel time-correlated single photon counting system (TCSPC).
A full description of our measurement scheme is given in the appendix.
}
\label{fig:setup}
\end{figure*}

The correct operation of Shor's algorithm is verified independently of the physical platform of the universal quantum computer on which it is run.  However, boson sampling is native to linear optical experiments, which allows us to exploit experimental methods and fundamental properties of linear optics to develop procedures that provide strong evidence that the system is functioning properly.  Firstly, we are interested in finding configurations of optical circuits that engender large-scale, ordered, photonic quantum interference, to produce a predictable structure in the probability distribution of possible detection events.  Fully reconfigurable circuits, capable of implementing any unitary transformation on optical modes, are realisable with arrays of beamsplitters and phase shifters \cite{Reck:1994dz}, which have been demonstrated on partially reconfigurable waveguide circuits \cite{Matthews:2009gi,Shadbolt:2011bw}.  With large scale single-photon and multi-photon interference verified for a predictable experiment, on a fully characterised circuit \cite{lo-sci-322-563}, a reasonable assumption is that quantum mechanics holds and the system maintains correct operation as the circuit is continuously reconfigured to implement a random unitary operation.  

Secondly, we determine that the most likely route to incorrect operation is the unwanted introduction of distinguishably between photons
\footnote{Errors equivalent to photon loss are heralded at the detection stage.},
which destroys quantum interference \cite{Hong:1987vi} and effectively pushes the matrix description of the optical circuit from one with complex entries, to a real valued matrix, where classical algorithms can efficiently and precisely approximate matrix permanents corresponding to the classical probability of individual outcomes \cite{Jerrum:2004vq}.  This regime is readily accessible experimentally, for example by introducing temporal delay between photons.  The opportunity then exists to tune between ideally indistinguishable (quantum) and perfectly distinguishable (classical) data,
and measure the change in a suitably constructed metric.
Additionally, we find a metric that discriminates among unitaries, based on the change in quantum versus classical statistics.
Our experimentally informed approach is different to a recently proposed test to confirm that boson sampling statistics are not drawn from an unbiased probability distribution \cite{Aaronson:2013wc}, which we also demonstrate, but which has the drawback that is it does not distinguish between quantum and classical statistics.

In these experiments, we observe and exploit a regular structure in the quantum probability distribution generated by a circuit of continuously coupled waveguides, which arises from a phenomenon related to boson bunching, which we term \emph{bosonic clouding}.  Here, photons are found to cluster in different but nearby modes, in a superposition around two separate locations.  This has been observed for systems of two photons propagating in continuously coupled waveguides known as photonic quantum walks \cite{Bromberg:2009ie, Peruzzo:2010tq, Meinecke:2013ib, Matthews:2013vc}.
Here, we observe that bosonic clouding persists for systems of 3, 4, and 5 photons propagating in continuously coupled waveguides, but is absent for circuits described by random unitary matrices, even for 3 photons.  We observe dissipation of the bosonic clouds when distinguishability is introduced among the photons. While our observation of this basic behaviour of particles is of fundamental interest, the boson clouds provide a unitary witness and a discriminator for operation between the quantum and classical regimes, without having to reconstruct parts of the probability distribution.

We experimentally implement our verification methods with 3, 4, and 5 photon ensembles propagating in arrays of 21 continuously coupled waveguides, producing Hilbert spaces of $>$50,000 dimensions.  The computational complexity of regularly structured probability distributions arising from such systems is not known, however for random unitary devices, strong evidence exists that efficient classical simulation is impossible \cite{Aaronson:2011tj}.  Turning our attention to photonic networks described by random unitary matrices, we observe probability distributions with little or no apparent structure for 3 photons propagating in 9 randomly connected optical modes, and we experimentally test the verification procedures that rule out sampling from a flat probability distribution \cite{Aaronson:2013wc}.

\begin{figure*}[t!]
\includegraphics[trim=0 0 0 0, clip, width=1\linewidth]{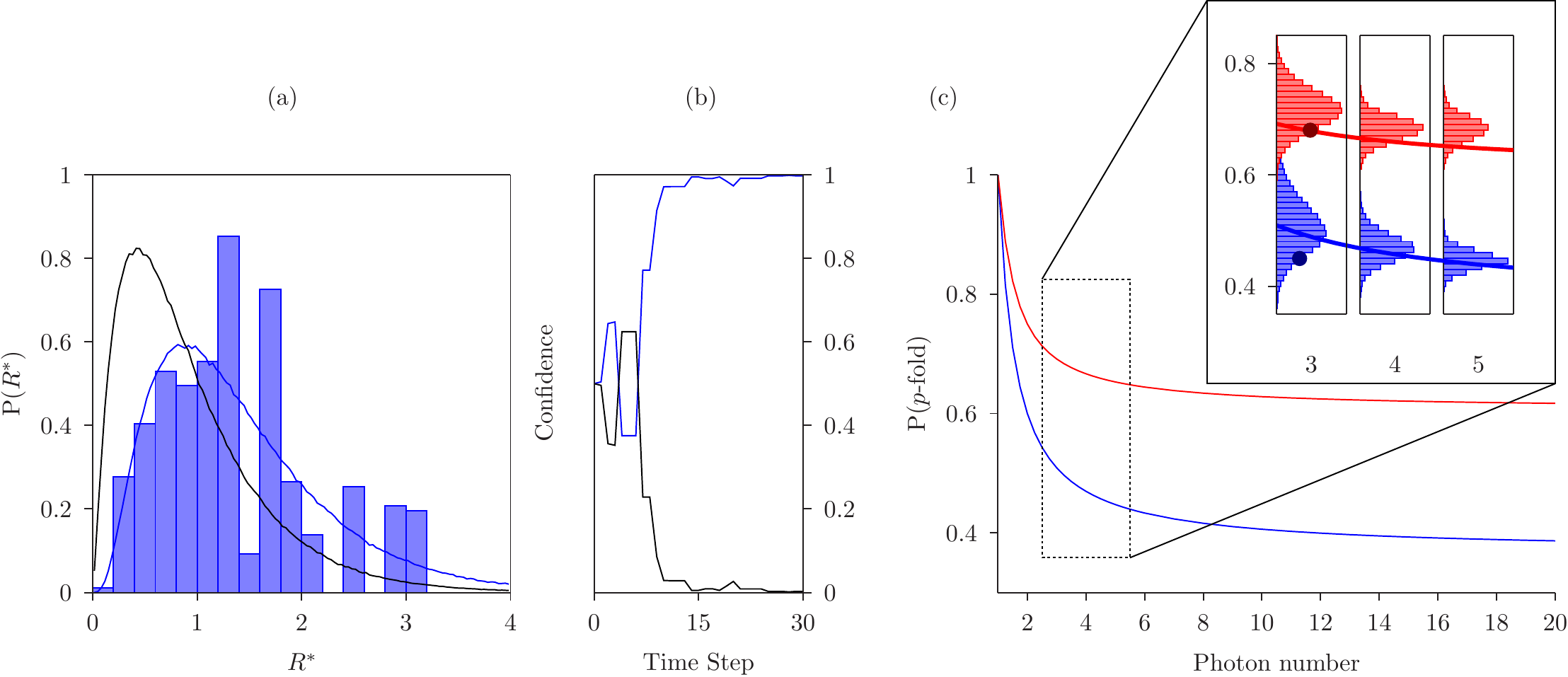}
\caption{
Verification of boson sampling $\dist{B}$ against the uniform distribution $\dist{F}$ (a,b) and discrimination between quantum $\dist{B}$ and classical $\dist{C}$ statistics (c).
(a) The expected probability density function for values of \(R^{*}\), for sub-matrices chosen from the boson sampling distribution (blue line) using the RU chip, and the uniform distribution (black line).
The bars show a histogram of \(R^{*}\) values from experimental 3 photon data.
(b) Dynamic updating using Bayesian inference for confidence in sampling from boson sampling distribution, rather than the uniform distribution.
After only 12 three-fold detection events we are over 90\% confident, and by the end of our experiment we assign only \(10^{-35}\) probability to the null hypothesis. 
(c) Probability of registering a \(p\)-fold click with both quantum (blue) and classical (red) particles.
The lines are asymptotic values with the constraint $m=p^2$, and the histograms (inset) are for numerical data up to 5 photons in 25 modes.
Values calculated for experimental data are shown over the inset histograms for 3 photons in 9 modes.}
\label{fig:device_independent}
\end{figure*}

All experiments presented here use a similar setup, displayed in Figure~\ref{fig:setup}, where two pairs of identical 780nm photons are generated from a pulsed spontaneous parametric downconversion source and injected into one of two waveguide circuits, which we label as QW and RU. The QW chip is a planar array of $21$ evanescently coupled single mode waveguides fabricated in silicon oxynitride ($\text{SiO}_{x}\text{N}_{y}$), with a circuit configuration similar to that used previously for photonic quantum walks of two photons.  The RU chip is a $9$ mode array of directional couplers and fixed phase shifts in silicon nitride ($\text{Si}_{2}\text{N}_{3}$), which can be fabricated to implement any fixed unitary operation \cite{Reck:1994dz}; here we have chosen a $9\times9$ Haar random unitary matrix.  Detection is performed over a simultaneous maximum of 16 modes with 16 single photon avalanche diodes (SPADs) and a 16-channel time-correlated single photon counting (TCSPC) system to monitor all $\binom{16}{p}$ $p$-fold events in realtime.

Our first experimental demonstration is motivated by the claim that boson sampling with an optical network described by a random unitary matrix is operationally indistinguishable from the case where detection events are drawn from an unbiased or flat probability distribution, with discrimination between the two only becoming possible after an exponential number of trials \cite{Gogolin:2013uh}.  We experimentally implement a procedure that, reasonably, uses knowledge of the unitary operation to efficiently verify that detection statistics are not collected from a flat probability distribution \cite{Aaronson:2013wc}. (Note that, even if the unitary description is a priori unknown, it can be efficiently measured, e.g. \cite{Laing:2012uw}).  The theoretical discriminator \(R^*\) is the product of row norms of the $p \times p$ sub-matrix $M$ that describes a transformation of $p$ photons, and is calculated from the complex matrix elements $\{a_{i,j}\}$ by computing, for each row, $R_i=|a_{i,1}|^2+|a_{i,2}|^2+\dotsb+|a_{i,p}|^2$, then taking the product $R^*=\prod_i^p R_i$ and normalising so that $E[R^*]=1$. Intuitively, this discriminator works because $R^*$ is sufficiently correlated with $|\textrm{Per}(M)|^2$, the probability of a $p$-fold detection given by the mod square permanent of the transformation sub-matrix.

\begin{figure*}[t!]
		\includegraphics[width=1\linewidth]{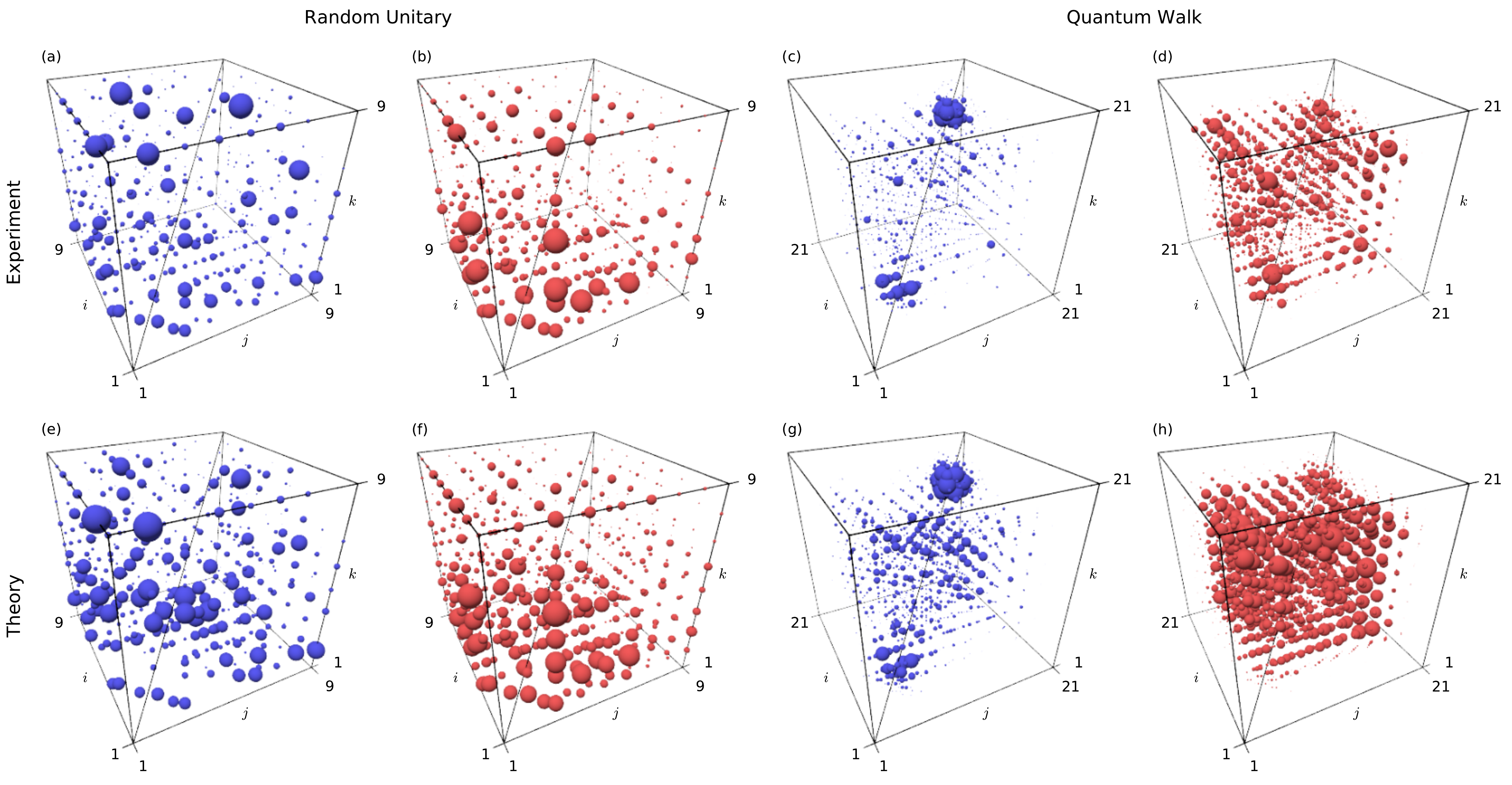}
	\caption{The absence and emergence of correlated bosonic clouds in three photon correlation cubes for a nine mode random unitary (a,b,e,f) and a 21 mode quantum walk (c,d,g,h).  The radii of spheres centred at coordinates ($i,j,k$) are proportional to the probability of finding three photons in output modes $i$, $j$ and $k$ respectively.  We tune between indistinguishable (blue) and distinguishable (red) photons by introducing a time delay between them.  These data are: (a) Experimental nine mode random unitary with indistinguishable and (b) distinguishable photons.  (c) Bosonic clouds from experimental 21 mode quantum walk unitary with indistinguishable and (d) distinguishable photons. (e)  Theoretical nine mode random unitary with indistinguishable and (f) distinguishable photons.  (g) Theoretical bosonic clouds from 21 mode quantum walk unitary with indistinguishable and (h) distinguishable photons.
The experimental data (top row) has been corrected for detector efficiencies and the theory has been filtered to show only events that were experimentally measured, which is the main reason for the apparent asymmetry between the pair of boson clouds.
	}
	\label{fig:cubes}
\end{figure*}

We collected 434 three-fold detections after injecting $p=3$ photon ensembles into our $m=9$ mode RU chip, shown in Fig.~\ref{fig:setup}(c), the unitary matrix description for which was reconstructed from single photon and two photon tomography \cite{Laing:2012uw}.  Figure~\ref{fig:device_independent}(a) shows a histogram of $R^*$ for these 434 events, together with numerical plots for both bosonic ($\dist{B}$) and flat ($\dist{F})$ distributions obtained by averaging over $10^5$ Haar random unitaries.  To quantify the performance of this discriminator, we use Bayesian inference to update in real time our confidence that the samples were drawn from $\dist{B}$ rather than $\dist{F}$.  Figure~\ref{fig:device_independent}(b) shows that after only 12 three-fold detection events a confidence level of $90\%$ that sampling is not from $\dist{F}$ is achieved, which rises to $1 - 10^{-35}$ by the end of the experiment.  See the appendix for details of this calculation.

A more physically relevant probability distribution to rule out, which is likely to be classically simulatable \cite{Aaronson:2013wc}, is that which is generated when photons become distinguishable, which we label as $\dist{C}$.  While $R^*$ discriminates between $\dist{B}$ and $\dist{F}$, it does not discriminate between $\dist{B}$ and $\dist{C}$.  Indistinguishability among photons may be verified at source \cite{Hong:1987vi}, yet the circuit may introduce distinguishability through decoherence, dispersion and other extra unwanted degrees of freedom such as polarisation.  We therefore implement a scalable method to verify that photon indistinguishability is maintained during propagation through the circuit, based around the question: given a $p$-photon input state in $p$ modes (one photon per mode) what is the probability, $P(p\textrm{-fold})$, of a $p$-fold detection?

\begin{figure*}[t!]
\includegraphics[width=\linewidth]{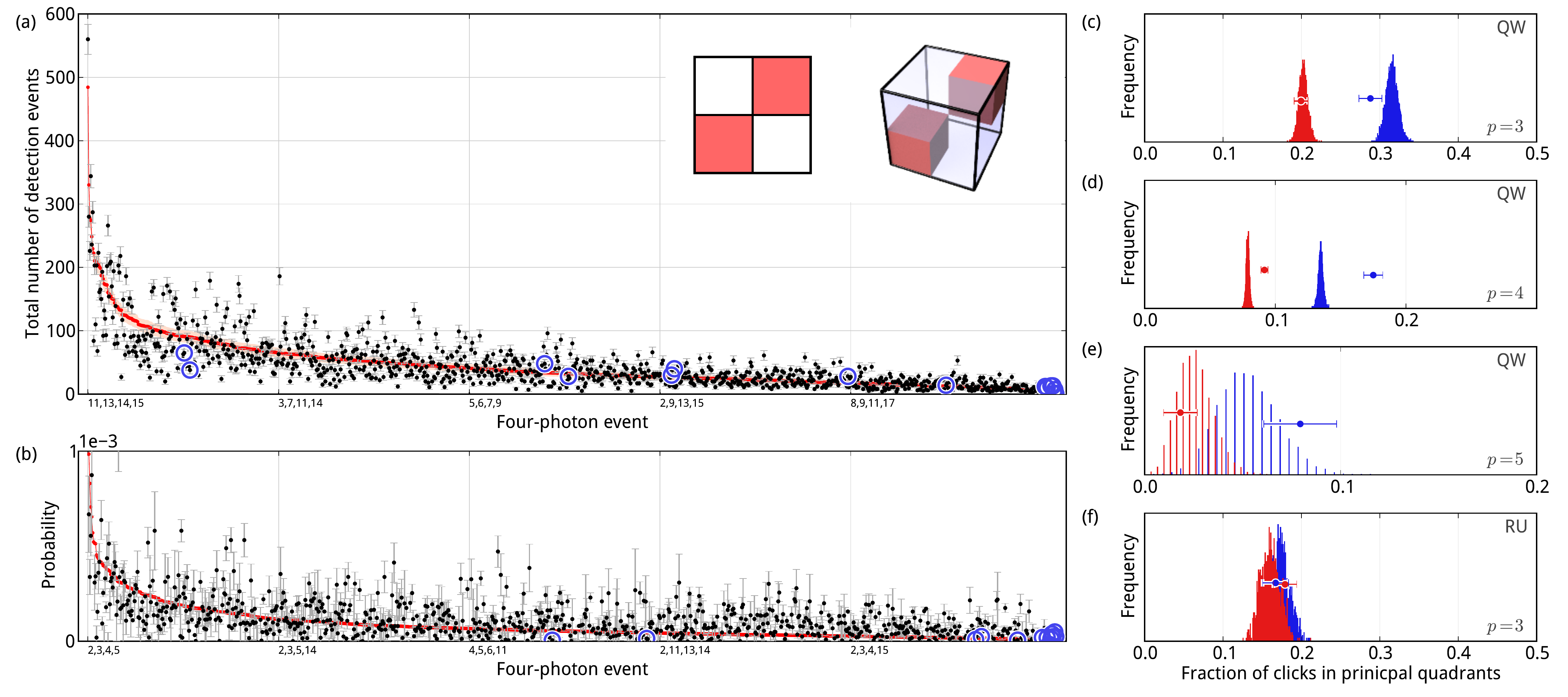}
\caption{Quantum-walk-specific verification. (a) Experimental data (black points) for four indistinguishable photons in a 21 mode quantum walk, over 1820 four-fold detection patterns, ordered by descending theoretical probability (red points). Number-resolved data is highlighted with blue circles.
Error bars assume Poissonian statistics. (b) Reconstructed pure-state four-photon data, after subtraction of experimentally-measured contributions due to $\ket{2200}$ and $\ket{0022}$ terms (see Appendix for details).
In (c-e) we perform a quantum-walk-specific test for $p= 3,4,5 $ photons, measuring the fraction of events $C$ in the principal quadrants (see inset). We plot experimental results for indistinguishable (blue) and distinguishable (red) photons, along with a corresponding theoretical distribution with the same number of samples drawn. In all cases, we see a statistically significant increase in $C$ for indistinguishable photons. In (f) we perform the same test for three photons in a 9-mode random unitary, where our quantum-walk-specific test does not reveal statistically significant quantum-classical separation, as expected.
}
\label{fig:unitary_specific}
\end{figure*}

The intuition, that $p$-fold detection is \emph{less} likely for \emph{indistinguishable} photons due to bosonic bunching, is formalised in \cite{Arkhipov:2011wt} with a simple counting argument to show that, when averaged over the Haar measure, in the case of indistinguishable photons $P^Q(p\textrm{-fold})=  \binom{m}{p} \Big/ \binom{m+p-1}{p}$ while in the case of distinguishable photons, as in the classical \emph{birthday paradox} $P^C(p\textrm{-fold}) = \binom{m}{p}p! \Big/ m^p$.  If $m \gg p^2$, $P^Q(p\textrm{-fold}) \approx P^C(p\textrm{-fold})$, however if $m=p^2$ the equivalence does not emerge, as can be seen in Figure~\ref{fig:device_independent}(c) where a separation emerges between $P^Q(p\textrm{-fold})$ and $P^C(p\textrm{-fold})$ \cite{Spagnolo:2013vr}.

The protocol requires $N$ trials of $p$-photon input states, which gives rise to $M$ $p$-fold detector events, allowing the comparison $M/N$ to the analytic values of $P^Q(p\textrm{-fold})$ and $P^C(p\textrm{-fold})$.  Due to the non deterministic nature of the downconversion process, we employ the method of \cite{Spagnolo:2013vr} to calculate $P(p\textrm{-fold})$ .  We found  $P^Q(p\textrm{-fold}) = 0.450\pm0.028$ compared to an expected value $0.509$, while $P^C(p\textrm{-fold})=0.680\pm0.0002$ compared to an expected value of $0.691$. Using the numerically determined probability density functions (pdf), shown in Fig.~\ref{fig:device_independent}(c), we estimate the probability (over Haar unitaries) that quantum data is the result of distinguishable particles to be $3\times 10^{-3}$, while the probability that classical data is the result of indistinguishable particles is $2\times 10^{-2}$.

Taken together, the tests in Fig.~\ref{fig:device_independent} provide circumstantial evidence that a boson sampling machine is operating according to the laws of quantum mechanics \cite{Aaronson:2013wc}, with non-trivial dependence on circuit parameters, and exhibiting quantum interference.  However, we now present a method which gives even stronger evidence for correct operation.  Consider implementing a highly structured unitary that promotes all of the essential physical features of boson sampling, including single photon and large scale multi-photon interference, but where significantly large parts of the probability distribution of $p$-fold detector clicks can be determined efficiently, classically, without calculating permanents of random matrices.  After experimentally confirming the structured probability distribution, the optical circuit can be tuned (continuously) to realise a unitary operation, such as a Haar random unitary, that produces a classically inaccessible probability distribution. The reasonable assumption is that correct operation is maintained during this process. Such a system could be realised by highly reconfigurable circuitry~\cite{Reck:1994dz,Shadbolt:2011bw}.  In our proof of principle experimental demonstration --- at the classically tractable scale where correct sampling from both probability distributions can be verified --- we physically swap between circuits.

The \emph{structured} unitary operation we choose is our QW chip of continuously coupled waveguides \cite{Bromberg:2009ie}.  The effect of this unitary operation is to produce bosonic clouding behaviour with photons observed to cluster, in superposition, around two separate local groups of modes.  We injected three photons into the middle ($k=10,11,12$) waveguides of the 21 mode QW chip (Fig.~\ref{fig:setup}(c)) and measure 524 out of a possible 1771 three-photon events, using fibre splitters and multiple detectors to achieve nondeterministic number resolved photon detection.

Experimentally obtained probability distributions for the QW chip, for both indistinguishable and distinguishable (temporally delayed) photons are compared with theoretical models in Fig.~\ref{fig:cubes}(c,d,g,h).  We found a statistical fidelity $F=\sum_{i}{\sqrt{p^{exp}_{i}p^{th}_{i}}}$  between the normalised theoretical $p_i^{th}$  and experimental $p_i^{exp}$ probability distributions of $F_Q= 0.930 \pm 0.003$ and $F_C= 0.961 \pm 0.002$ for the indistinguishable and distinguishable case respectively.  Errors bars are calculated by propagating Poissonian count rate errors and the deviation from unit fidelity can be attributed to the presence of higher order terms, and the temporal distinguishability between non-pair photons \cite{Tanida:2012eh}. 
For the theory model, by assuming a nearest neighbour Hamiltonian \cite{Perets:2008gb}, single photon tomography is sufficient to yield the unitary description of the circuit.  

Bosonic clouding behaviour can be clearly seen for indistinguishable photons (Fig.~\ref{fig:cubes}(c)), that is photons cluster around the main diagonal line of the correlation cube, where probabilities exactly on this line correspond to full bunching of all three photons in the same mode.  Two clouds have formed at separate locations in the cube centred on modes 6 and 16; this means if one photon is detected in the locality of mode 16 (for example), the remaining two photons have a high probability of being correlated to this event and also detected around mode 16. In contrast, when temporal distinguishability is introduced between all photons (Fig.~\ref{fig:cubes}(d)), quantum interference is destroyed and the clouds dissipate: there is now a higher probability that the two remaining photons will be found away from the modes local to mode 16. 

For further comparison, we have also presented all 84 possible three-photon correlated detection probabilities in the RU chip in Fig.~\ref{fig:cubes}(a,b,e,f). We note that the clouds observed in Fig.~\ref{fig:cubes}(c) are absent in the RU chip for both indistinguishable Fig.~\ref{fig:cubes}(a) and distinguishable photons Fig.~\ref{fig:cubes}(b); the correlation cubes do not reveal any discernible structure.  We found a fidelity between our experiment and theoretical model of $F_Q=0.939\pm0.010$ and $F_C=0.970\pm0.007$, for indistinguishable and distinguishable photons, respectively .

To use the bosonic clouding behaviour as a QW witness we construct a metric which determines how many events lie in the clouded regions by dividing the correlation hypercube in half along each axis, which for $p$ photons creates \(2^{p}\) quadrants
\footnote{Technically, this geometrical region is termed an orthant or hyper-octant}.
(see the inset of Fig.~\ref{fig:unitary_specific}(a)), and calculating the fraction of events $C$ which occupy the two principle quadrants
In the case of perfect clouding all photons are output from the same half of the interferometer, meaning all events are in the principle quadrants and $C=1$.  If there are always at least two photons detected in different halves of the interferometer, then $C=0$, which means zero events in the principle quadrant.  To include cases of approximate clouding, where most photons are localised, one can consider clusters of quadrants around the principle quadrant, but we expect this to be the subject of future investigations and do not consider them here.

Numerical simulations confirm that $C$ can witness the correct operation of QW unitaries after a small number of trials for up to $p=7$ photons, which is supported by our experimental evidence for up to $p=5$ photons.  For the three photon case we find $C=0.288 \pm 0.015$, compared to a numerically determined value of $0.332 \pm 0.008$.  For $p=4$ photons we use a novel sifting technique (see appendix for details) to yield statistics for four photons injected into the central ($k=9,10,11,12$) waveguides (one photon per mode) and we measure 1016 out of a possible 10626 four-fold events as shown in Fig.~\ref{fig:unitary_specific}(a,b).
The fidelity between experimental and theoretical probability distributions was found to be $F_Q = 0.971 \pm 0.001$ and $F_C = 0.978 \pm 0.0004$.  For the case of $p=4$ photons we measured $C=0.175 \pm 0.007$ compared to $0.144 \pm 0.002$.  For $p=5$ photons we observe 217 5-fold detection events from the four mode input state, corresponding to a six photon creation event where one photon is lost, and we found $C=0.079 \pm 0.019$ compared to $0.058 \pm 0.016$. These values deviate from our numerical simulations because of experimental imperfections (e.g. photon distinguishability and partial mixture for $p=5$), however by tuning between the distinguishable and indistinguishable case, we observe a change in $C$ that witnesses the formation of bosonic clouds.  

This change is depicted in Fig.~\ref{fig:unitary_specific}(c,d,e) and for three photons was found to be $\Delta C=0.089\pm0.017$ compared to $0.130\pm 0.009$, for four photons was $\Delta C = 0.083 \pm 0.008$ compared to $0.065 \pm 0.002$, and for five photons was $\Delta C = 0.061 \pm 0.020$ compared to $0.033 \pm 0.018 $.  In the five photon case, even though the state is partially mixed, bosonic clouding can still be observed with only a small number of experimental samples in a vast ($>50,000$ dimensional) Hilbert space.  For comparison, in Fig.~\ref{fig:unitary_specific}(f) we plot $C$ for the random unitary, and find $\Delta C = -0.012 \pm 0.022$, confirming the absence of clouding.

Here we have shown how to combine physical phenomena associated to a physical system with experimental and technological capabilities in that system, to provide evidence of correct operation for quantum algorithms that may be formally unverifiable.  In doing so, we demonstrate bosonic clouding which is interesting in its own right.  However, each platform for quantum technologies will exhibit its own unique features, for example anti-bunching due to the Pauli exclusion principle.  We expect \emph{machine level} verification techniques to continue to be important and, increasingly, techniques will need to keep pace with the growing scale and complexity of quantum systems \cite{lu-arxiv-2013}.

During the final revision of this manuscript, the efficient experimental validation of photonic boson sampling against the uniform distribution has been reported online \cite{sp-arxiv-2013}.

\begin{acknowledgements}
The authors would like to thank G.\ Marshall, E.\ Mart\'{i}n L\'{o}pez,  A.\ Peruzzo, A.\ Politi and A.\ Rubenok for technical assistance.
\end{acknowledgements}

\clearpage

\appendix
\section{BAYESIAN INFERENCE}
We use Bayesian inference to update in real time our confidence that the device is not sampling from the uniform distribution, using the row-norm product estimator \(R^*\). As shown in Figure~\ref{fig:device_independent}(a), the pdf for this variable is different depending on whether samples are drawn from the flat distribution \(\dist{F}\) or the boson sampling distribution \(\dist{B}\), with a high value of \(R^*\) being more likely in the latter case.

We have determined numerically (by averaging over $10^5$ random unitaries chosen by the Haar measure ) that for the case of \(p=3\) photons and \(m=9\) modes:
\begin{align*}
  P \left( \left(R^* > 1\right) | \dist{B} \right) &= 0.631,\\
  P \left( \left(R^* < 1\right) | \dist{B} \right) &= 0.369,\\
  P \left( \left(R^* > 1\right) | \dist{F} \right) &= 0.355,\\
  P \left( \left(R^* < 1\right) | \dist{F} \right) &= 0.645 .
\end{align*}
Given these probabilities we can use the value of \(R^*\) computed from a detection event to update our confidence that the device is sampling from \(\dist{B}\) rather than \(\dist{F}\) according to Bayes' theorem
\begin{equation*}
  P \left( \dist{H} | R^{*} \right) = \frac{P \left( R^{*} | \dist{H} \right) P \left( \dist{H} \right)}{P \left( R^{*} \right)},
\end{equation*}
where for \(\dist{H}\) we substitute either \(\dist{B}\) or \(\dist{F}\). Our prior is \( P \left( \dist{B} \right) = P \left( \dist{F} \right) = 0.5 \).

The reason for choosing 1 as a threshold for \( R^{*} \) is that this was the method proved to be a scalable discriminator in \cite{Aaronson:2013wc}. Although the numerically determined pdf in Figure~\ref{fig:device_independent}(a) for boson sampling events cannot be computed in a scalable way, the pdf for uniform sampling can. In fact, in the limit of large photon number \(p\) and large mode number \(m \gg p^{2}\) it tends to lognormal, so \(P \left( \left(R^* > 1\right) | \dist{F} \right)\) and \(P \left( \left(R^* < 1\right) | \dist{F} \right)\) can be computed. The result in \cite{Aaronson:2013wc}, that
\begin{equation*}
  P \left( \left(R^{*} > 1\right) | \dist{B} \right) - P \left( \left(R^{*} > 1\right) | \dist{F} \right) \geq \frac{1}{9}
\end{equation*}
can then give us bounds on \(P \left( \left(R^* > 1\right) | \dist{B} \right)\) and \(P \left( \left(R^* < 1\right) | \dist{B} \right)\). We are not operating in the limits where this applies, so computing the values numerically is more appropriate.

\section{DEVICE DETAILS} 
\label{sec:devices}

\subsection{9 mode unitary} 
\label{sub:9_mode_unitary}
The $m=9$ mode random unitary was fabricated in silicon nitride ($\text{Si}_{2}\text{N}_{3}$) with a refractive index contrast $\Delta = (n_{\textrm{core}}^2 - n_{\textrm{cladding}}^2)/2n_\textrm{core}^2=27\%$.  The waveguides had a width of $1.5\mu\text{m}$ and outside the interaction region were separated by $127\mu\text{m}$.   The device consisted of 36 directional couplers whereby the waveguides are bought to within $2.5\mu\text{m}$ of one another, for an interaction length of $\sim400\mu\text{m}$ (dependent on the desired splitting ratio).  The fibre to fibre coupling efficiency was $\sim5\%$.

\subsection{21 mode quantum walk unitary} 
\label{sub:21_mode_quantum_walk_unitary}
The $m=21$ waveguide array was fabricated in silicon oxynitride ($\text{SiO}_{x}\text{N}_{y}$). The index contrast of $2.4\%$ enables fabrication of micron sized single mode waveguides in compact circuit designs with a minimum bend radius of $560\mu\text{m}$. The waveguides are designed with a constant width of $2.2\mu\text{m}$ and height of $0.85\mu\text{m}$. They are pitched at $1.3\mu\text{m}$ within the coupling region of length $700\mu\text{m}$ in order to achieve sufficient mode overlap for nearest neighbour coupling. The waveguides bend adiabatically to a pitch of $127\mu\text{m}$ at the input and output facets to match the standard separation of the fibre arrays we butt couple to the chip. The waveguides are tapered to a width $0.7\mu\text{m}$ at the facet to achieve better mode overlap with the fibre modes, this way we obtain an overall fibre to fibre coupling efficiency of  $\sim 30\%$.


\section{DETECTION SCHEME} 
\label{sub:detection_scheme}

Generally, off-diagonal elements in the $p$-photon correlation matrix (i.e.\ the collision free subspace where $j_1\neq j_2\neq j_3...\neq j_p$) can be measured without the use of fibre splitters, with waveguide outputs connected directly to APDs. Diagonal elements, where $j_a=j_b$ for some $a,b \in \left[ 1,\ldots, p \right], a \neq b$ give the probability of finding two or more photons in the same spatial mode and must be measured using fibre splitters and multiple detectors to achieve nondeterministic number resolved photon detection.

In particular for the three photon case correlations of the form $\ket{2_i1_j}$ and $\ket{1_i1_j1_k}$ were measured between all even numbered and all odd numbered waveguides.  Further we measured all possible three photon coincidences, including $\ket{3_i}$, across the four sets of waveguides $\{1-5\}$,$\{6-10\}$,$\{11-15\}$, and $\{16-20\}$.  In total we measured 524 out of a possible 1771 three fold events.

For the four photon case we measured collision free events of the form $\ket{1_i1_j1_k1_l}$ across the set of 16 waveguides numbered $\{3-10,12-19\}$ and all possible four photon coincidences across waveguides $\{3,7,11,15\}$.  We measured 1016 out of a possible 10626 four photon events.


\section{FOUR PHOTON INPUT STATE} 
\label{sub:1111_input_state}

\begin{figure}[h]
\centering
\includegraphics[width=0.3\textwidth]{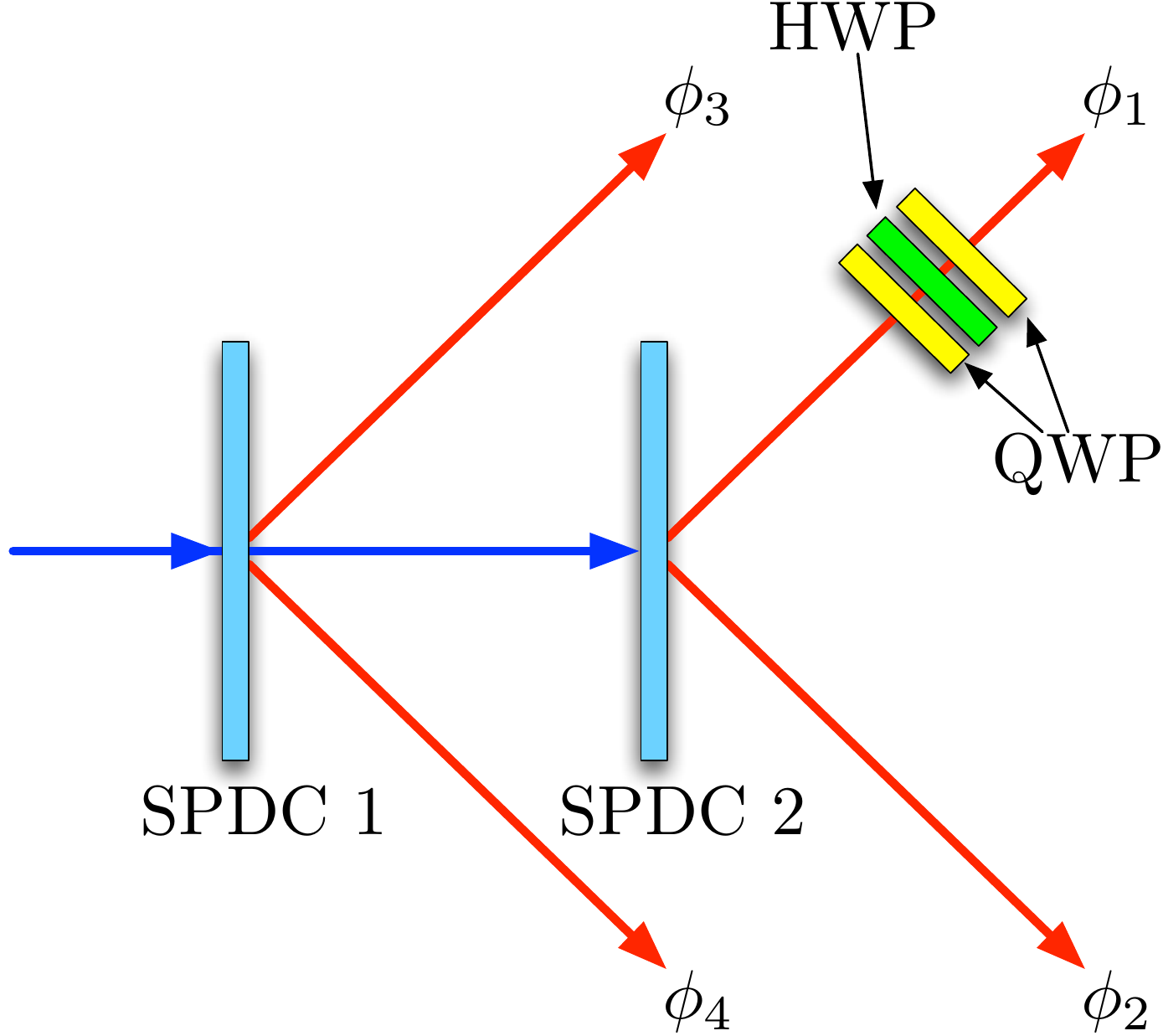}
\caption{A four mode two crystal down-conversion system with unknown phases $\phi_i$ in the $i^\textrm{th}$ mode, and a half (HWP) and two quarter waveplates (QWP) to force the system into a maximal mixture.}
\label{fig: spdc}
\end{figure}

The challenge in yielding four photon data for the non-superposed $\ket{\psi}_4=\ket{1111}$ input state, is isolating this state from the entire down-conversion state.  We do this in postselection by applying a series of phases to one arm of the down-conversion, which when summated yield a maximal mixture.  We then subtract the superposition terms separately to yield statistics for $\ket{\psi}_4$.  To see this consider the entire four photon subspace of the two crystal, four mode down-conversion state
\begin{equation*}
\begin{split}
\label{eq: SPDC}
\ket{\psi}_{\textrm{SPDC}}= & \frac{1}{\sqrt{3}}\left(e^{i(\phi_1+\phi_2+\phi_3+\phi_4)}\ket{1111} \right. \\
                           & \left. +e^{2i(\phi_3+\phi_4)}\ket{0022}+e^{2i(\phi_1+\phi_2)}\ket{2200} \right),
\end{split}
\end{equation*}
where $\phi_i$ is an unknown phase $\phi$ in the $i^\textrm{th}$ mode of the down-conversion as in Figure~\ref{fig: spdc}.

We apply a phase $\theta$ to mode 1 by inserting half (H) and quarter (Q) waveplates in the configuration $\textrm{Q}(\pi/4)\textrm{H}(\theta/4+\pi/4)\textrm{Q}(\pi/4)$, to give a state $\rho_{\theta}$.  It can be shown that
\begin{equation*}
	\begin{aligned}
		\rho_0+\rho_{\pi/2}+\rho_{\pi}+\rho_{3\pi/2}=&\ket{1111}\bra{1111}+\ket{0022}\bra{0022} \\
																								&+\ket{2200}\bra{0022} \\
																								=&\rho_\textrm{mix}
	\end{aligned}
\end{equation*}
where $\rho_\textrm{mix}$ is the four photon maximally mixed state.  By inputting modes 1,2 and 3,4 we yield data for $\rho_{2200}=\ket{2200}\bra{2200}$ and $\rho_{0022}=\ket{0022}\bra{0022}$, which can be subtracted from $\rho_\textrm{mix}$ to give $\ket{\psi}_4$.

The phases $\phi_i$ are a function of time. However, our scheme yields statistics for the mixed state independent of the speed of phase fluctuation. The only necessary assumption is that on average the phases $\phi_i$ are not correlated with the phase $\theta$.

Note that this is not intended to be a scalable solution to isolating large states like $\ket{11...1}$.



\end{document}